# ACDC-Tracing: Towards Anonymous Citizen-Driven Contact Tracing

*Kristof Roomp, Microsoft* kristofr@microsoft.com
*Nuria Oliver, Spanish Royal Academy of Engineering & ELLIS Unit Alicante,* nuria@alum.mit.edu

## Executive Summary

As we enter the control phase of the COVID-19 pandemic, many efforts have been dedicated to developing smartphone-based contact tracing apps in order to automatically identify people that a person with COVID-19 might have infected. These applications while potentially useful, present significant adoption, societal, technical and privacy challenges.

We propose ACDC-Tracing, a simpler, anonymous, voucher-based contact tracing solution that relies on peoples' knowledge of their own close contacts. People who test positive are given an anonymous voucher which they can share with a limited number of people whom they think they might be infected. The recipients can use this voucher to book a COVID-19 test and can receive their test results without ever revealing their identity. People receiving positive result are given vouchers to further backtrack the path of infection.

This is a fully anonymous solution which does not require any sharing of location data, Bluetooth, or having an app installed on people's mobile device. Moreover, ACDC-Tracing can be tested for effectiveness at a small scale without requiring adoption by the entire population, which would enable acquiring fast evidence about its efficacy and scalability. Finally, it is compatible with and complementary to alternative approaches to contact tracing.

## Introduction

As we enter the control phase of the COVID-19 pandemic, *early detection* of a significant portion of all infected individuals will be of critical importance, so that they can be placed in quarantine and thus avoid further contagion. To achieve this, several key infrastructures need to be in place, including (1) the ability to perform daily massive amounts of tests, (2) a person-centric end-to-end protocol for isolating infected or vulnerable individuals, (3) special measures for individuals in high-risk, high-spread professions, such as healthcare personnel and nursing home workers; and (4) the ability to give tests to the right people to quickly identify new sources of infection.

## Contact Tracing via Smartphone Apps

It is regarding this last point --the detection of infected individuals and their contacts—where the concept of *contact tracing via smartphone apps* arises. To date, several app proposals with varying levels of privacy implications have been made to track contacts in order to quickly find possible infections when someone tests positive for COVID-19. Unfortunately, most of these proposals – and even the most privacy-preserving ones, such as the SafePaths project at MIT[1] and the PEPP-PT project in Europe[2] – have several limitations, including:

- **Critical mass**: To be effective, it is estimated that at least 60% of the population[3] must be running the application. To make it even more challenging, in Europe only 76% of the population has smartphones capable of running these apps[4]. This is as high as 82% in the UK and as low as 58% in Italy[5], with prominent socio-economic and demographic differences. For example, younger children, who can spread the illness without having symptoms, generally do not have smartphones.
- **Privacy and potential misuse**: These apps require citizens to record very personal data, which may make people reluctant to run the app once they realize the potential consequences of misuse (even in the case when data is only stored locally on their phone). For example, could law enforcement oblige someone to unlock their phone and analyze it to determine all the people they have been in contact with? In a domestic violence or divorce situation, could someone deliberately misreport, or threaten to misreport, an infection on their partner's phone? Once this app is present on everyone's phone, will governments be tempted to use contact tracing for other purposes? When receiving a notification from their phone, will people try to figure out who they got it from, potentially starting malicious rumors?

- **Cybersecurity**: Many of these mobile apps are being developed very quickly by new, relatively immature software development teams, which could result in serious security vulnerabilities[6]. Magnifying the problem, since the same software would be used across an entire country, this would present an extremely high-value target for both state-based and non-state-based bad actors[7].
- **Accuracy**: Phone-based urban geolocation is only accurate to about 10 meters[8], and research using Bluetooth LE for proximity tracing has been mostly done in controlled environments with known hardware[9]. Accuracy is especially important in apartment buildings where many people are close by measured distance, but not necessarily sharing the same physical space. The result of these inaccuracies could generate a very large set of people to test, creating an immense demand for testing with many false positives.

As part of the insights[10] from the analysis of the answers to the Covid19ImpactSurvey, we found that a high proportion (over 70%) of people that tested positive could identify their likely source of infection (e.g. a member of their household, a relative, a friend, a patient, client or co-worker). This finding made us consider whether a much simpler solution could consist of simply asking people whom they might have infected and scale this process to work with a large number of people.

## ACDC-Tracing

The proposed solution, called Anonymous Citizen-Driven Contact Tracing (ACDC-Tracing), would work as follows:

1. Person X receives a positive diagnosis for COVID-19. As part of receiving this diagnosis, Person X is given a "COVID-19 testing voucher", which consists of a short alpha-numeric code. This voucher could be provided either via SMS, e-mail, a randomly selected sealed envelope or whatever other mechanism makes sense depending on the testing protocol of the country. This voucher is not connected to the identity of Person X.
2. Person X is told that they can share this code with up to 6 people, which is an upper bound of the estimated reproductive number for COVID-19[11]. This number would be adjusted based on the findings of an initial pilot. Person X would be asked to distribute these vouchers to the people they think they might have infected.
3. The recipients of the code can use it to book an appointment for receiving a COVID-19 test at a web site provided by the government or by a private testing facility. The web site does not require the recipient to identify themselves; it only validates that the code has not been used more than 6 times and gives a confirmation code with a testing location and a date/time.
4. The recipient goes to the testing facility, presents the confirmation code and is given the test. The next day, the recipient can check the results of their test either e.g. online or by phone using only their confirmation code (which would be fully anonymous). Recipients that test positive would again receive a voucher that would allow them to follow the same process as above.

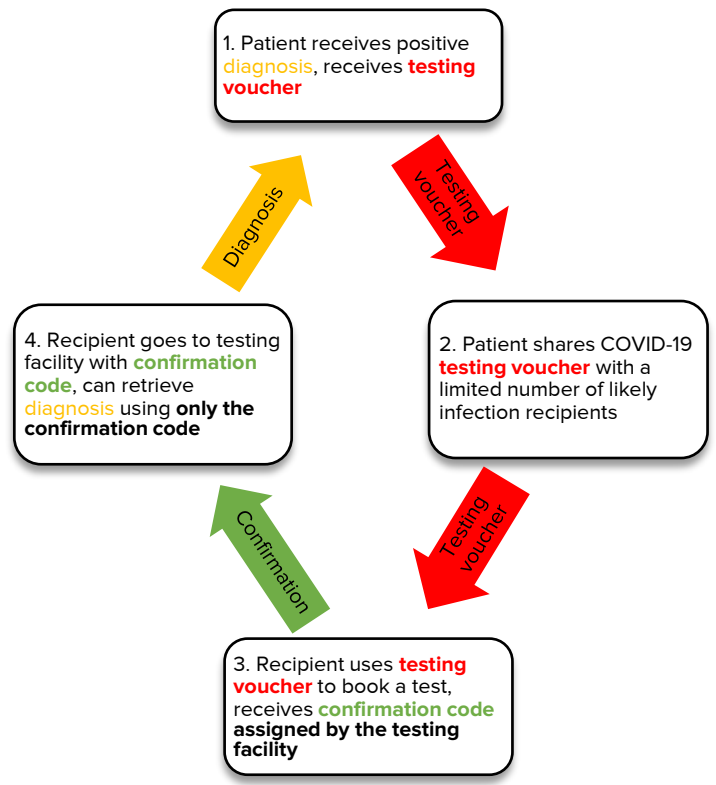

## No Critical Mass Required to be Effective

Phone based contact tracing proposals require a large uptake by the general population before being effective, and do not prove their value until a critical mass is reached.

A key advantage of ACDC-Tracing is that it can be evaluated and tested at a small scale to prove its effectiveness.

## Privacy

There are several key points concerning privacy preservation in this proposal:

1. Vouchers are not tied to identity. The only thing that needs to be tracked is the number of times that a voucher has been used. Once this is done, all information related to that voucher can be erased.

2. No trusted 3rd party involvement is required in the communication between the infected person and the possible recipients. Person X can send the voucher via their preferred means of communication, by phone, by an encrypted WhatsApp conversation, etc.
3. No personal data is stored on any mobile device --or anywhere for that matter-- that could be compromised by security breaches.

## Integration with phone-based contact tracing

ACDC-Tracing is complementary to other contact tracing approaches and could be integrated into a phone-based solution. For example, instead of just notifying people that they have been in close contact with an infected individual, a contact-tracing app could provide an anonymous single-use voucher that would allow people to immediately get tested, using the same infrastructure already created for ACDC-Testing.

## Discussion

The proposed approach is scalable, technologically simple and fully privacy preserving, while at the same time taking advantage of a "viral" model for tracing back the infection chain. We believe that it could be implemented in a short time with a simple back-end infrastructure. It could even be implemented without any infrastructure using paper vouchers in sealed envelopes (with a matching checklist at the testing center). This could be especially important in areas where there is high risk of stigmatization due to a positive diagnosis, since no names or identification need to be recorded.

While technology is a key element to help tackle the spread of COVID-19, important further considerations need to be taken into account for any early detection plus contact tracing approach, such as: (1) the availability of testing infrastructure for a large number of people; (2) the existence of a person-centric quarantine where people are not afraid to test positive; (3) the empowerment of and trust placed in citizens who would be the main source of the contact tracing; and (4) the development of incentives for asymptomatic, low-risk individuals to get tested if they were at risk of having contracted COVID-19.

ACDC-Tracing is compatible with other contact tracing options. It could provide a fast, high-quality solution in a timely fashion. In addition, it allows us to quickly learn which types of contact tracing are most effective, providing valuable ground truth to develop models that identify which contacts are most likely to produce positive test results.